\documentclass[
superscriptaddress,
 amsmath,amssymb,
 aps,
floatfix,
]{revtex4-2}

\usepackage{graphicx}
\usepackage{dcolumn}
\usepackage{bm}

\usepackage[usenames, dvipsnames]{color}
\makeatletter
\let\cl@chapter\undefined
\makeatletter
\usepackage[capitalise]{cleveref}
\usepackage{soul}

\graphicspath{{figures/}{../figures/}}

\newcommand{\ba}{\begin{eqnarray}}
\newcommand{\ea}{\end{eqnarray}}
\newcommand{\be}{\begin{equation}}
\newcommand{\ee}{\end{equation}}
\newcommand{\benn}{\begin{equation*}}
\newcommand{\eenn}{\end{equation*}}
\newcommand{\eps}{\epsilon}
\begin{document}

\preprint{}

\title{Separatrix crossing and symmetry breaking in NLSE-like systems\\ due to forcing and damping}

\author{D. Eeltink}
\affiliation{Group of Applied Physics and Institute for Environmental Sciences, University of Geneva, Switzerland }%
\author{A. Armaroli}%
\affiliation{Group of Applied Physics and Institute for Environmental Sciences, University of Geneva, Switzerland }
\author{C. Luneau}
\affiliation{Institut Pytheas, AMU,CNRS,IRD Marseille, France}%
\author{M. Brunetti}%
\affiliation{Group of Applied Physics and Institute for Environmental Sciences, University of Geneva, Switzerland }%
\author{H. Branger}
\affiliation{IRPHE, AMU,CNRS,ECM Marseille, France}%
\author{J. Kasparian}%
 \email{jerome.kasparian@unige.ch}
\affiliation{Group of Applied Physics and Institute for Environmental Sciences, University of Geneva, Switzerland }%

\date{\today}

\begin{abstract}
We theoretically and experimentally examine the effect of forcing and damping on systems that can be described by the nonlinear Schr\"odinger equation (NLSE), by making use of the phase-space predictions of the three-wave truncation of the spectrum. In the latter, only the fundamental frequency and the upper and lower sidebands are retained. Plane wave solutions to the NLSE exhibit modulation instability (MI) within a frequency band determined by a linear stability analysis. For modulation frequencies inside the MI-band, we experimentally demonstrate that forcing and damping cause a separatrix crossing during the evolution. Our experiments are performed on deep water waves, which are better described by the higher-order NLSE, the Dysthe equation. We therefore extend our analysis to this system. However, our conclusions are general. When the system is damped by the viscosity of the water, it is pulled outside the separatrix, which in the real space corresponds to a phase-shift of the envelope and therefore doubles the period of the Fermi-Pasta-Ulam-Tsingou recurrence cycle. When the system is forced by the wind, it is pulled inside the separatrix. Furthermore, for modulation frequencies outside the conventional MI-band, we experimentally demonstrate that contrary to the linear prediction, we do observe a growth and decay cycle of the plane-wave modulation. Finally, we give a theoretical demonstration that forcing the NLSE system can induce symmetry breaking during the evolution.
\end{abstract}

\maketitle

\section{Introduction}\label{sec_Intro}
The Nonlinear Schr{\"o}dinger equation (NLSE) describes the propagation of the field-envelope in many different systems, for instance in optical fibers, Bose-Einstein condensates, water waves, and Langmuir waves in hot plasmas \cite{Dudley2014,Pitavskii2001,Kimmoun2016,Fried1973}. Elementary solutions of the NLSE include plane waves, solitons and breathers. The plane wave solution is subject to modulation instability (MI)~\cite{Zakharov1968}: the linear stability analysis of the NLSE reveals that within a certain frequency bandwidth, a modulation - perturbation - to the plane wave will grow exponentially. It therefore modulates the amplitude of the plane wave, generating a train of sharp pulses~\cite{Wetzel2011}. Remarkably, the MI can exhibit cyclic behavior, known as the Fermi-Pasta-Ulam-Tsingou (FPUT) recurrence \cite{Akhmediev2001}: despite complex nonlinear dynamics, the system returns to its initial condition.

We are interested in the effect of forcing and damping on the dynamics of the system, specifically on the recurrence cycle. Many systems that can be described by the NLSE naturally undergo dissipation, however, not many allow to be forced \cite{Akhmediev2005}. Water waves can undergo both: while viscosity is a natural source of damping, wind can provide forcing. 

Thus we performed experiments in a wind-wave facility to corroborate our theoretical results. To accurately describe the asymmetries in the spectrum of water waves, the higher-order version of the NLSE, the Dysthe equation \cite{Dysthe1979} is required. Unlike the NLSE, there are no known analytic solutions to the Dysthe equation. 

To study the essential physical behavior of the NLSE, and allow explicit calculations, the spectrum can be truncated to only three-wave components: the main mode, and upper and lower sidebands \cite{Trillo1991}. A phase-space can be spanned by the relative amplitudes of sidebands with respect to the main mode $\eta_\textrm{F}$ and the relative phase $\psi$ between the sidebands and the main mode. In the same manner, a three-wave truncation can be performed for the Dysthe equation \cite{Armaroli2017a}.

While the primary effect of forcing and damping is to make the amplitude grow and decrease, respectively, their influence on the phase-space is nontrivial. In this paper, the three-wave truncation allows us to trace the trajectory of the wave tank measurement in the phase plane,  offering an explicit understanding of the complex evolution of the system. For modulation frequencies inside the MI-band, we experimentally demonstrate that dissipation attracts trajectories outside the separatrix, whereas forcing attracts them to the inside. As such, forcing and damping can cause a separatrix crossing during the evolution of the system \cite{Cary1986,Bourland1990}. Furthermore, while no modulation is expected outside of the MI-band, we experimentally demonstrate the growth and decay cycle of solutions as predicted by \cite{Conforti2019} in this regime. In addition, we  perform long distance simulations in which forcing induces a symmetry breaking in the three-wave phase-space: by moving the modulation frequency inside the MI-band, the Hamiltonian is transformed from a single to a double well.

\section{Theory}
\begin{figure*}
\centering
\includegraphics[width=0.99\textwidth]{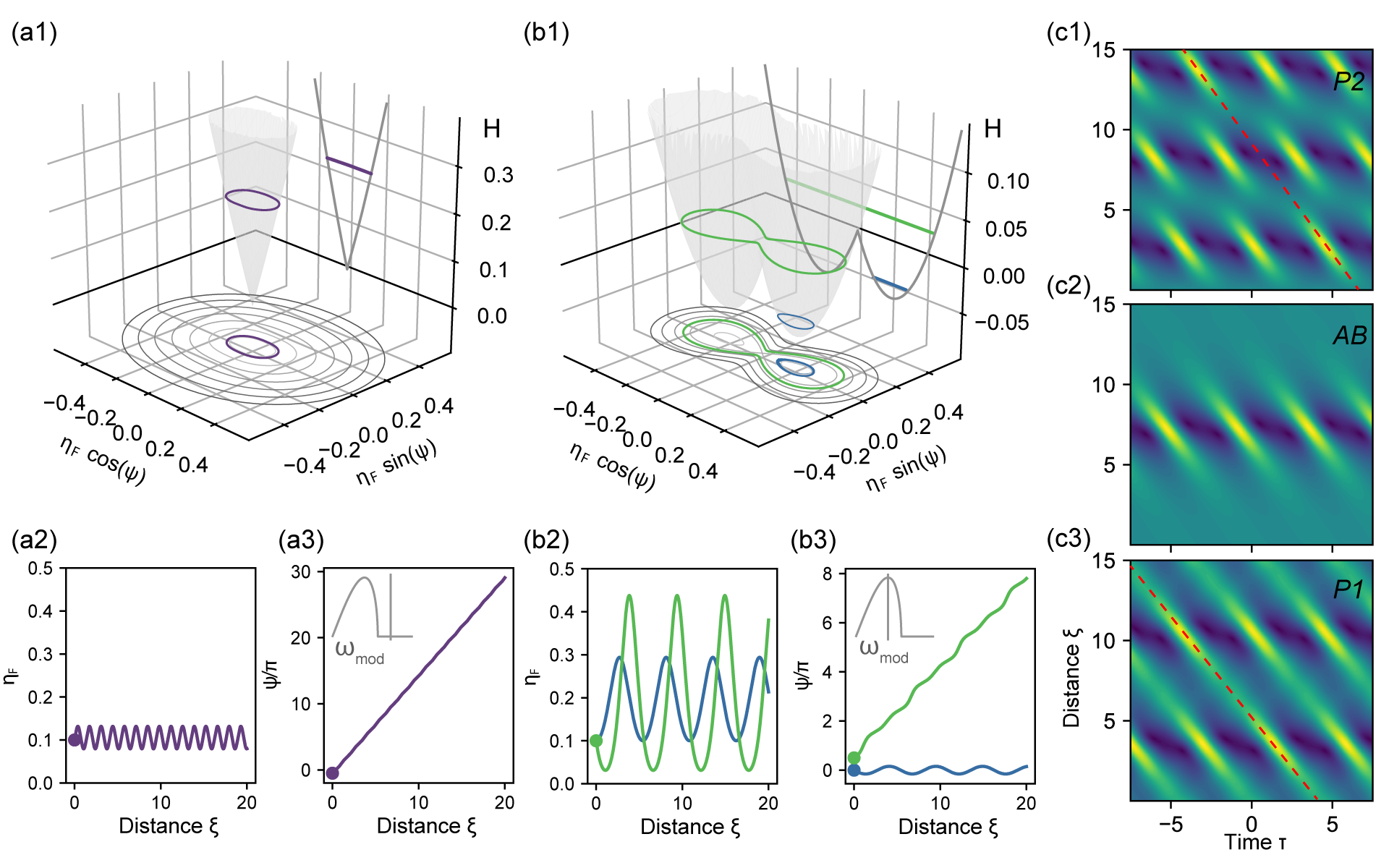}
\caption{Conservative dynamics of the three-wave model based on the Dysthe equation without the $2 a^2\frac{\partial a^*}{\partial \tau}$ term. (a1,b1) Wrapped phase-space of $\eta_F$ and $\psi$. Initial conditions: $\epsilon=0.05$, $\alpha=0$. (a) $\Omega_\text{mod}=2.4$: outside of the MI-band. Purple: $\psi=0$, $\delta_0=\delta_1=0$. (b) $\Omega_\text{mod}=\sqrt{2}$:  inside the MI-band, $\delta_0=\delta_1=0$. Green:~ $\psi=\pi/2$. Blue: $\psi=0$. See also Supplementary movie 1. (a2,b2) Corresponding evolution of $\eta_\textrm{F}$. (a3,b3) Corresponding evolution of $\psi$. (c) Evolution of the envelope in the Dysthe equation \cref{eq:Dysthe:FDadim}. (c1) P2 solution (green in b), (c2) AB with $\xi_0$ = -5 and (c3) P1 solution (blue in b). 
} \label{fig:PhaseSpace}
\end{figure*}

To take into account the viscous-damping and the wind-forcing of water waves, we developed the forced-damped Dysthe equation for the propagation of the envelope \cite{Eeltink2017}: 
\be\label{eq:Dysthe:FDadim}
\underbrace{ \frac{\partial a}{\partial \xi} + i\frac{1}{2}  \frac{\partial^2 a}{\partial \tau^2} + i a|a|^2}_\textrm{NLSE} =   \underbrace{\delta_0a + i\delta_1 \frac{\partial a}{\partial \tau}}_\textrm{Damping / Forcing}  
+\eps\underbrace{\bigg(8 |a|^2\frac{\partial a}{\partial \tau} + 2 a^2\frac{\partial a^*}{\partial \tau} + 2ia \mathcal{H}\left[\frac{\partial |a|^2}{\partial \tau}\right]\bigg)}_\textrm{Dysthe terms}\\ 
\ee
\noindent where $a$ is the envelope, $\xi$ is adimensional space, $\tau$ adimensional time, and $\delta_0,\delta_1$ the balance between forcing and damping at the leading and higher order, respectively. The steepness $\eps=A_0 k_0 \sqrt{2}$, where $A_0$ is the reference amplitude of an
ideally stable Stokes’ wave and $k_0$ the wavenumber. Quantities have been adimensionalized in the following way:
\begin{align}
 t' &= t-x/c_{\textrm{g, lin}} & a &= \tilde{a}/A_0 \\
    \tau &= t'/T_0, & T_0 &= 1/(\omega_0\eps)  \\
    \xi &= x/L_0, & L_0 &=1/(2\eps^2k_0)  \\
    \delta_0 &=\frac{T_0}{2\eps}\left(\Gamma - 4k_0^2\nu\right) & \delta_1 &=  2T_0\left(\Gamma - 5k_0^2\nu\right) 
\end{align}
\noindent where $c_{\textrm{g, lin}}=\frac{1}{2}\left(g/k\right)^{-1/2}$ is the linear group velocity in the deep-water limit, $\tilde{a}$ the dimensional envelope, $\nu$ the kinematic viscosity and $\Gamma$  the wind growth rate.

The forced-damped Dysthe equation gives good agreement with experiments thanks to the higher-order terms that reproduce the asymmetries in the spectrum \cite{Eeltink2017}. The three-wave truncation allows the system to be described more thoroughly as it enables to study its dynamics explicitly. Below we briefly recall the relevant notation and conclusions from an earlier work~\cite{Armaroli2018_3w}.

Assuming that the system behavior can be adequately described by restricting ourselves to three modes,  namely a main mode and two sidebands, we can write for the envelope $a$:
\be
a(\xi, \tau) = a_0(\xi) + a_1(\xi)e^{-i\Omega \tau}+ a_{-1}(\xi)e^{i\Omega \tau}
\ee
That is, the wave is reduced to a harmonically perturbed plane wave (HPPW). Inserting this into \cref{eq:Dysthe:FDadim} gives a system of three ordinary differential equations for  $\frac{\partial a_m}{\partial \xi}$ (Eq. (5) of \cite{Armaroli2018_3w}). Writing
\be
a_m(\xi) = \sqrt{\eta_m(\xi)}e^{i\phi_m(\xi)} \quad, \quad (m = 0, \pm1)
\ee
\noindent allows to construct a closed system of equations, consisting of an evolution equation in $\xi$ for each of the following quantities (Eq. (6) in \cite{Armaroli2018_3w}):
\begin{subequations}\label{eq_3w_def}
\begin{align}
&N_3&\equiv & \quad \eta_0 + \eta_1 + \eta_{-1} & \text{Norm} \label{eq_3w_def_Norm}  \\ 
&\eta_\text{F} &\equiv& \quad  (\eta_1 + \eta_{-1})/N_3 & \text{Sideband fraction} \label{sys_A_Norm} \\
& \psi  &\equiv&\quad  (\phi_1 + \phi_{-1}-2\phi_0)  & \text{Relative phase} \label{sys_A_NS}\\ 
& \alpha &\equiv &\quad (\eta_1 - \eta_{-1})/N_3 & \text{Sideband imbalance} \label{sys_A_DBC} 
\end{align}
\end{subequations}
\noindent This three-wave system closely describes the dynamics of the full spectrum, where the sideband imbalance $\alpha$ is the three-wave counterpart of the spectral mean, and $N_3$ that of the full norm.

\subsection{Conservative Dynamics}
For the three-wave system, a phase-space ($\eta_\text{F} \cos(\psi)$, $\eta_\text{F} \sin(\psi)$) can be constructed as in \cref{fig:PhaseSpace}. The level sets mark a constant Hamiltonian of the conservative Dysthe equation ($\delta_0=\delta_1=0$), neglecting the term $2 a^2\frac{\partial a^*}{\partial \tau}$ that is partly responsible for the growth of the spectral asymmetry \cite{Armaroli2017a}.  It is well known that the linear stability analysis of the NLSE reveals that a plane wave is unstable to perturbations with a modulation frequency $\Omega_\text{mod}<2$,  and a maximum instability at $\Omega_\text{mod}=\sqrt{2}$. For the Dysthe equation the MI-band is slightly modified, depending on the wave-steepness $\eps$ \cite{Dysthe1979}. Like for the conventional NLSE, see \cite{Mussot2018} for a comprehensive overview, the three-wave Hamiltonian has a single-well shape when $\Omega_\text{mod}$ is outside the MI-band (\cref{fig:PhaseSpace}a1). When $\Omega_\text{mod}$ is inside the MI-band, it is a double-welled (\cref{fig:PhaseSpace}b1). See supplementary movie 1 of this symmetry breaking as a function of ${\Omega_\text{mod}}$. 
\subsubsection{Outside the MI-band}
The oval shape of the Hamiltonian level sets in \cref{fig:PhaseSpace}a show that, contrary to the prediction of the linear stability analysis, there is an oscillation of the sideband amplitude $\eta_\text{F}$ (as exemplified by the purple trajectory in (a2)), and thus a growth and decay cycle of the envelope amplitude. 
\subsubsection{Inside the MI-band}
For $\Omega_\text{mod}$ inside the MI-band (\cref{fig:PhaseSpace}b,c), a separatrix marks the boundary in the double-welled landscape, separating the two types of trajectories in the phase plane. The separatrix corresponds to the Akhmediev Breather (AB) solution in the conservative NLSE frame. However, because the Dysthe equation is not integrable, the separatrix corresponds to the AB solution only for small initial sidebands and neglecting small fluctuation in the spectral mean.  

Outside the separatrix (i.e. higher Hamiltonian values), trajectories  (displayed in green in \cref{fig:PhaseSpace}b) undergo a $\pi$ phase shift at each recurrence of $\eta_\text{F}$, so that the period of the whole system is twice that of the $\eta_F$\footnote{We use the term `recurrence' however, strictly speaking, this is a quasi-recurrence, since in the Dysthe equation there is no an exact return to the initial conditions. This holds even stronger when the system is damped and forced. However, we use this term to refer to the general process of modulation-demodulation and oscillation of $\eta_F$ and $\psi$.}. We therefore term these trajectories period-2 or P2 solutions (green).  As the phase crosses $\psi =  n \pi$ (\Cref{fig:PhaseSpace}b3) a phase-shift occurs. \Cref{fig:PhaseSpace}c1 shows the real-space evolution of the solution in which the phase-shift is apparent as a shift of the second focal point with respect to the dashed line. 

For lower Hamiltonian values, closed trajectories remain within the separatrix, and have the same period for $\psi$ and $\eta_\textrm{F}$. We term these period-1 or P1 solutions (blue).
\subsubsection{Link to Type A and Type B solutions}\label{sec:Akhmediev}
Ref. \cite{Akhmediev1987_MI} derives a three parameter family of solutions of the NLSE, of which the Akhmediev breather (AB), Kuznetsov-Ma and Peregrine breather are special cases. In this framework, a phase-space can be spanned by $\mathcal{R}(a(\xi,\tau_m))$, $\mathcal{I}(a(\xi,\tau_m))$, where $\tau_m=n/\Omega_\text{mod}$ is the time-point where maximal modulation occurs, see for instance \cite{Soto-Crespo2017}.  

The AB forms the separatrix between two types of solutions, labelled type A and type B. Type A-solutions share the characteristic phase-shift with the three-wave P2 solutions, whereas type B solutions, like P1 solutions, show no phase shift, and therefore have a period half that of type A. Note however that in this phase-space, type A solutions are on the inside of the separatrix, and type B solutions are on the outside.

In addition, type A solutions can grow outside of the MI-band, whereas type B solutions do not \cite{Conforti2019}. Differently put, based on the single well in \cref{fig:PhaseSpace}a1, only P2 or type A solutions exist outside of the MI-band.

\subsection{Nonconservative dynamics: Damping and forcing}
Based on the above reminders, it is clear that in a conservative system ($\delta_0=\delta_1=0$), the initial condition starts in either the P1 or a P2-basin and stays there. P1 and P2 trajectories have been experimentally observed in fiber optics by choosing the corresponding initial value for $\psi$ \cite{Mussot2018,Naveau2019}.

For a non-conservative system, however, we deduce in \cite{Armaroli2018_3w} that the attraction basin is determined by the leading-order term of the forcing/damping balance $\delta_0$. In the viscous regime ($\delta_0<0$),  the solution is attracted to the P2 solution outside the separatrix. In the wind-forced regime ($\delta_0>0$), the solution is attracted to the P1 solution, {\it i.e.} to an evolution without phase-shift. Damping and forcing changes the norm $N_3$ and thus contracts / expands the phase space: as such, the separatrix can be crossed during the evolution.

Dissipation is naturally present in most systems. For dissipative water waves, long-tank experiments have demonstrated the phase-shifted P2 trajectories \cite{Kimmoun2016,Kimmoun2017}, and consequently the doubling of the FPUT recurrence frequency. In the present work we shall experimentally show that forcing attracts to P1 trajectories and that damping and forcing allows to cross the separatrix during the evolution.

\section{Experimental setup and simulation parameters}
The main goal of our experiment is to demonstrate that wind-forcing can attract trajectories towards P1 solutions. In doing so, we use forcing to cross the separatrix during the evolution. The attraction from one regime to another can take several recurrence cycles, depending on the initial distance to the separatrix. Therefore, the tank length is a critical limiting factor. While for dissipative experiments very long tanks (up to 250 meters \cite{Kimmoun2016,Kimmoun2017}) are available, typically closed air-loop wind-facilities are much shorter.  Furthermore, a certain amount of wind-forcing is needed to overcome the viscous dissipation. However, too much wind forcing induces wave breaking, which is a form of dissipation, and therefore sets an upper bound to the amount of forcing. The combination of these factors drastically narrows the window to observe the opposing behaviour of damping and forcing.

In addition, we examine the behavior for an initial condition outside the MI-band, for which we confirm the prediction that there are indeed growth and decay cycles in the damped case ({\it i.e.} the experiment is performed without wind). 

\subsection{Experimental setup}

\begin{figure*}
\centering
\includegraphics[width=0.7\textwidth]{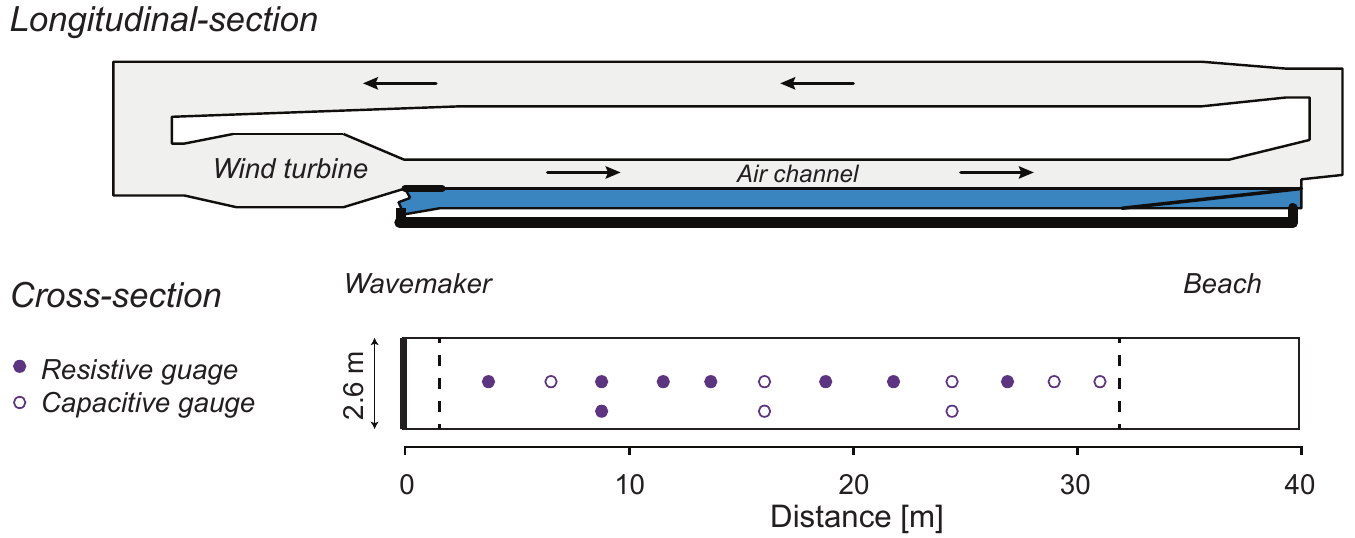}
\caption{Wind-wave facility: side- and top-view, not to scale. The tank has a water depth of 80 cm. An 8 m sloping beach prevents wave reflection. At the beginning of the tank, a 1.5 m long floating sheet damps possible high-frequency mechanical wave modes and guides the incoming wind tangential to the water surface. The air-channel above the tank is 1.5 m high. 12 wave gauges were placed in the center of the tank, and 3 wave gauges were placed 30 cm from the side-wall.
} \label{fig:waveTank}
\end{figure*} 

Experiments have been performed in the 40 m long closed wind-wave facility at IRPHE/PYTHEAS (Luminy) Aix Marseille University, see \cref{fig:waveTank} for details on the dimensions. Mechanical waves have been generated by an underwater piston wave maker. The system was able to produce arbitrary surface gravity waves in the frequency range up to 1.9 Hz. The wind was generated by a closed-loop air flow system, in the direction of the wave propagation, blowing continuously. A total of 15 wave gauges were used, of which 12 were placed approximately evenly in the center of the tank (central wave gauges), and 3 were placed off-center (transverse wave gauges) to account for transverse waves.  All gauges had a sampling rate of 400 Hz.

\begin{table*}
\centering
\caption{Experimental and simulation parameters. $\Gamma$ in $10^{-3}$  1/s, $\nu$ in $10^{-6}$ m/s$^2$. }
\label{tab:parameters}      
\begin{tabular}{cccccccccc}
\hline\noalign{\smallskip}
 Wind [m/s] &  $\Gamma$ Sim. & $\nu$ Exp.& $\nu$ Sim. & $\delta_0$ Sim. & $\Omega_\text{mod}$& $f_0$ & $\eps_0$ & Experiment  \\
\noalign{\smallskip}\hline\noalign{\smallskip}
\multicolumn{8}{l}{Initial condition inside MI-band: Akhmediev breather} \\
\hline
0      & 0    & 2     & 2  &  -0.007  & $\sqrt{2}$ & 1.70 & 0.12 & \checkmark \\
   3.1     & 6.0      & -     & 2  &  0.032  & $\sqrt{2}$ & 1.70 & 0.12 & \checkmark \\
      4.0     &  -      & -     & -  & -  & $\sqrt{2}$ & 1.70 & 0.12 & \checkmark \\
   \hline
\multicolumn{8}{l}{Initial condition outside MI-band: Harmonically perturbed plane wave} \\
\hline
0      & 0    & 2     & 2  &  -0.006  & $3$ & 1.35 & 0.10&\checkmark  \\
0     & 0   & 2     & 2  &  -0.006  & $2.4$ & 1.35 & 0.10&\checkmark \\
-     & -   & -     & -  &  0.05  & $2.4$ & 1.35 & 0.10&x \\
\noalign{\smallskip}\hline
\end{tabular}
\end{table*}

\subsection{Initial condition inside the MI-band}

The initial condition has to be close to the separatrix, to allow the transition of the solution from P2 to P1 within the available tank length. We therefore initialized the wave maker with a signal reproducing the AB:
\be \label{eq_AB}
a(\tau, \xi) =
\frac{\sqrt{2A} \cos{\Omega_\text{mod} \tau} + (1-4A)\cosh{2 R\xi} + i R \sinh{2R\xi}}{\sqrt{2A} \cos{\Omega_\text{mod} \tau} - \cosh{2R\xi}}
\ee
\noindent where $\Omega_\text{mod} = 2\sqrt{1-2 A}$ and $R=\sqrt{8A(1-2A^2)}$. In the Dysthe-based three-wave phase-space, this corresponds to starting slightly on the outside of the separatrix.

To get a maximal dimensionless distance, the carrier wave frequency was chosen close to the upper limit of the wavemaker at 1.70 Hz. To avoid wave breaking (see Section \ref{sec:Results}) we limit our background steepness to $\eps = 0.12$.  In addition, we started as close as possible to the focal point without having too much deformation of the initial condition: we used $x_f =- 8$ m ($\xi = -1.8 $) in \cref{eq_AB} to have the focal point after 8 meters of propagation. This wave-train was launched in different wind conditions, with wind blowing at a continuous speed of 0, 3.1 and 4.0 m/s. 

Simulations were performed based on the complex envelope extracted from the measurement of the first wave gauge as initial condition. The viscosity $\nu$ was set to a fixed value of $2 \times 10^{-6}$~m/s$^2$ for all simulations, based on the dissipation value we calculated for the runs without wind,  see \cref{tab:parameters}. This value is higher than the theoretical value  $1 \times 10^{-6}$~m/s$^2$ in order to account for the damping due to the sidewalls. The wind input parameter $\Gamma$ can be theoretically calculated using the Miles mechanism \cite{Miles1957,Conte1959} to be $4.8 \times 10^{-3} \textrm{s}^{-1}$. As this is only an estimation, we tuned this parameter such that it matched what we observed in experiments, see \cref{tab:parameters}.
\begin{figure*}
\centering
\includegraphics[width=0.90\textwidth]{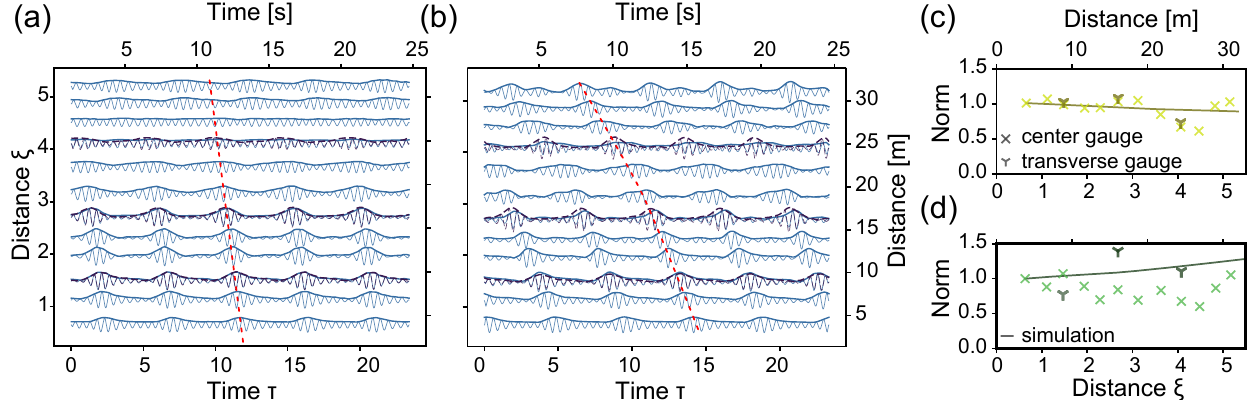}
\caption{(a,b) Surface elevation and envelope (thick line) in the frame of the linear group velocity, offset by wave gauge position for (a) no wind (b) wind speed of 3.1 m/s. Dashed lines correspond to transverse modes. (c,d) Norm $\int |a|^2 dt$ from simulations and measurements for (c) no wind (d) wind speed of 3.1 m/s} \label{fig:Waterfall}
\end{figure*} 

\subsection{Initial condition outside the MI-band}
To examine the behavior outside the MI-band, our initial condition was a plane wave seeded by an upper and lower sideband, with sideband fraction $\eta_F = 0.05$, unbalance $\alpha=0$, and relative phase $\psi=-\pi/4$. This HPPW was launched for two different modulation frequencies: $\Omega_{\text{mod}} = 2.4$ (\cref{fig:outsideMI}a)  and $\Omega_{\text{mod}} =3$. As the sidebands are further away from the main mode, a lower carrier wave frequency was used than for the AB: $f_0$ = 1.35 Hz., to keep the higher modes within reach of the wavemaker.
Experiments were performed without wind. For the simulations where the system is forced, the three-wave model \cite{Armaroli2018_3w} was integrated with the same initial conditions as the experiment.

\subsection{Phase extraction}
The phase information is crucial to trace out the trajectory on the phase plane and distinguish P1 from P2 trajectories (or Type A from Type B as in \cite{Conforti2019}). However, to our knowledge no publications exist that show the experimental phase evolution of the complex envelope. While the complex envelope can be reconstructed from the real-valued surface elevation using the Hilbert transform \cite{Osborne2010}, in order to create the phase space spanned by $\mathcal{R}(a(\xi,\tau_m)), \mathcal{I}(a(\xi,\tau_m))$, one has to track exactly $\tau_m$ which will evolve with the nonlinear group velocity. The noise of the Hilbert transform combined with the uncertainty of the exact group velocity give unreliable results.

We manage to experimentally obtain the phase information with the help of the three-wave truncation, as $\eta_F$ and $\psi$ are properties that can be obtained directly from the complex spectrum of the surface elevation, see \cref{eq_3w_def}b,c. Performing a fast Fourier transform on the latter yields the phase $\phi_m$ and amplitude $\eta_m$ of each mode. This eliminates both the computation of the complex envelope and the calculation of the group velocity which are too noisy to provide reliable results.

\section{\label{sec:Results}Results and Discussion}
\begin{figure}
\centering
\includegraphics[width=0.3\textwidth]{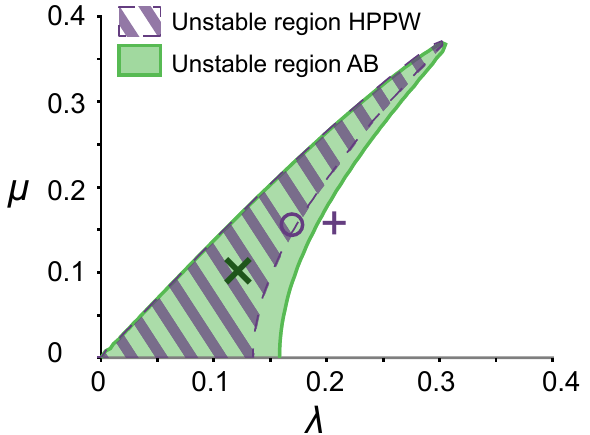}
\caption{Instability region of the BMNLS equation \cite{Trulsen1997}, for transverse ($\mu$) and longitudinal wavenumbers ($\lambda$). Green area: unstable region for the AB with $f_0=1.70$ Hz, $\eps=0.12$. Green cross: initial condition $\Omega_\text{mod}=\sqrt{2}$. Purple area gives the unstable region for the HPPW of $f_0=1.35$ Hz, $\eps=0.1$, where $\Omega_{\textrm{mod}} = 2.4$ (circle) and $\Omega_{\textrm{mod}} =3$ (cross) both lie outside the longitudinal MI-band.} \label{fig:transverseInstab}
\end{figure} 

\Cref{fig:Waterfall}a,b shows the measured surface elevation (thin lines) and envelope (thick lines) in the frame moving at the linear group velocity, and offset by the wave gauge distance, without wind (a), and for a wind speed of 3.1~m/s (b). The envelope is obtained through the Hilbert transform, neglecting the bound modes. The dark dashed lines are the measurements of the transverse wave gauges. The group velocity (dashed red line) increases and the modulation is amplified with wind.

\begin{figure*}
\centering
\includegraphics[width=0.7\textwidth]{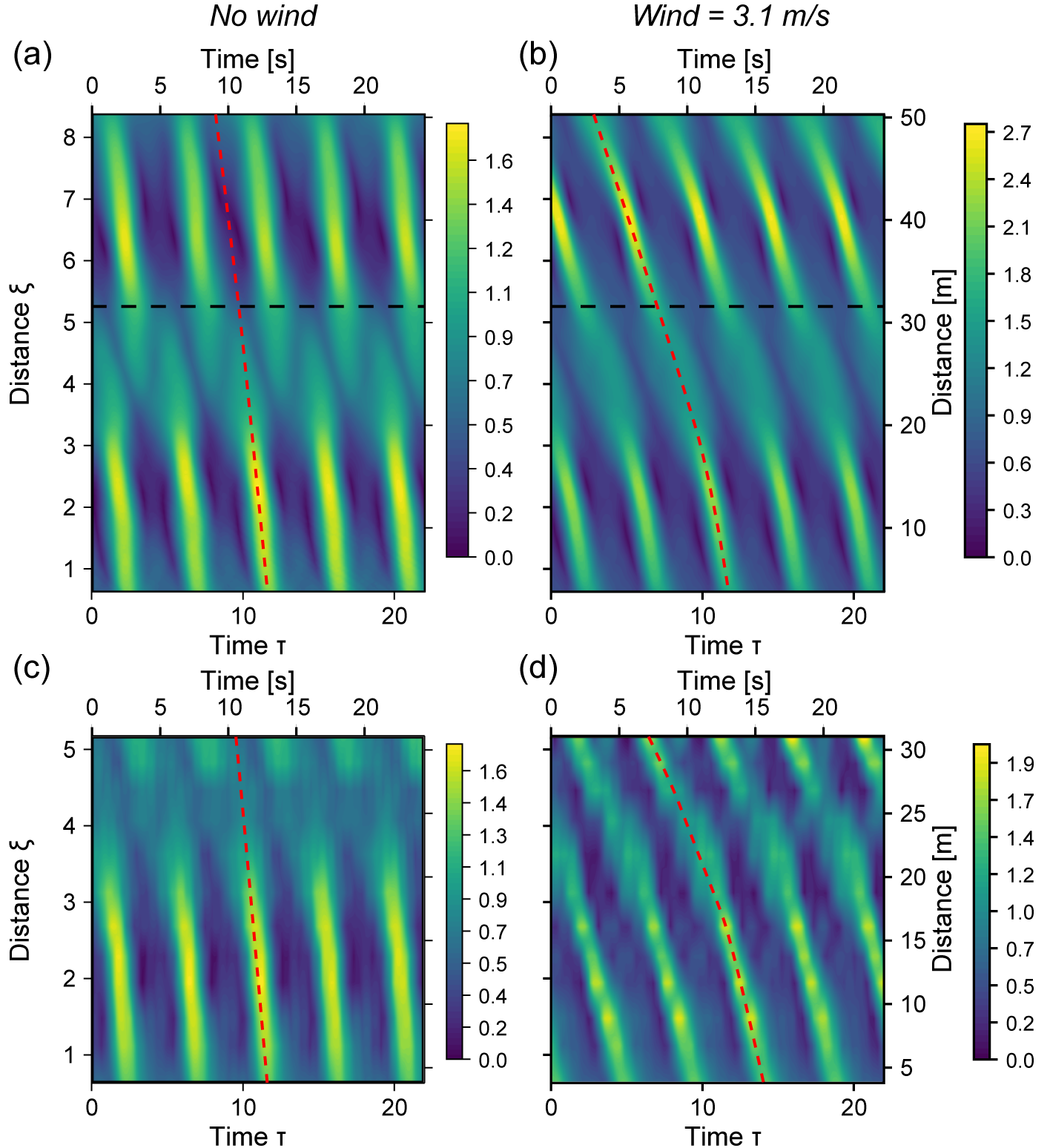}
\caption{Evolution of the envelope $|a|$ in $\xi$. The horizontal dashed line indicates physical tank limit. a) Simulation: Wind = 0 m/s, b) Simulation: Wind = 3.1 m/s,  c) Experiment: Wind = 0 m/s, d) Experiment: Wind = 3.1 m/s. } \label{fig:envelope}
\end{figure*} 

\Cref{fig:Waterfall}c,d shows the measured and simulated evolution of the norm.  The fluctuations in the experimental norm can be attributed to the transverse modes excited in the relatively wide wave tank. \Cref{fig:transverseInstab}, shows the instability region defined by \cite{Trulsen1997,Trulsen1999}. Due to the tank width $b$, there is an unstable band of the adimensional modulation in the transverse direction $\mu = \frac{k_\perp}{k_0}$, where $k_\perp = \pi / b$, as the longest transverse standing wave has a wavelength of $2b$.  The longitudinal modulation wavenumber $\lambda$ is based on the linear stability analysis of the broader bandwidth modified NLSE \cite{Trulsen1996}. For the AB, transverse modes (green cross in \Cref{fig:transverseInstab}) are predicted, that cannot be accounted for by our one-dimensional model. As the central-gauges lie on the nodes of the transverse modes, the most reliable source of the evolution of the norm are the transverse gauges. Only the latter are therefore used to estimate the growth and decay rates for the simulations (solid line in \cref{fig:Waterfall}c,d). Without wind, the norm decreases (\cref{fig:Waterfall}c); when the wind is blowing at 3.1 m/s the norm increases (\cref{fig:Waterfall}d).

The evolution of the envelope in the frame of the linear group velocity is displayed in \cref{fig:envelope}, in the forced and damped regimes, respectively. Panels a,b show the simulations,  where the limit of the tank-length is indicated by the black-dashed line. Panels c,d show the corresponding experimental measurements. Without wind, \textit{i.e.} when the system is damped, the envelope shifts in phase, indicated by a shift from the red dashed line for the second (quasi) recurrence cycle, both for the simulation (\cref{fig:envelope}a) and the measurement (\cref{fig:envelope}c). This indicates a P2 solution. When wind is blowing at 3.1 m/s, the system is forced, and the phase shift disappears: the modulation-crest of the second quasi-recurrence cycle is in line with that of the first, \textit{i.e.} the crest follows the red-dashed line, indicating a P1 solution. 

\begin{figure*}
\centering
\includegraphics[width=0.7\textwidth]{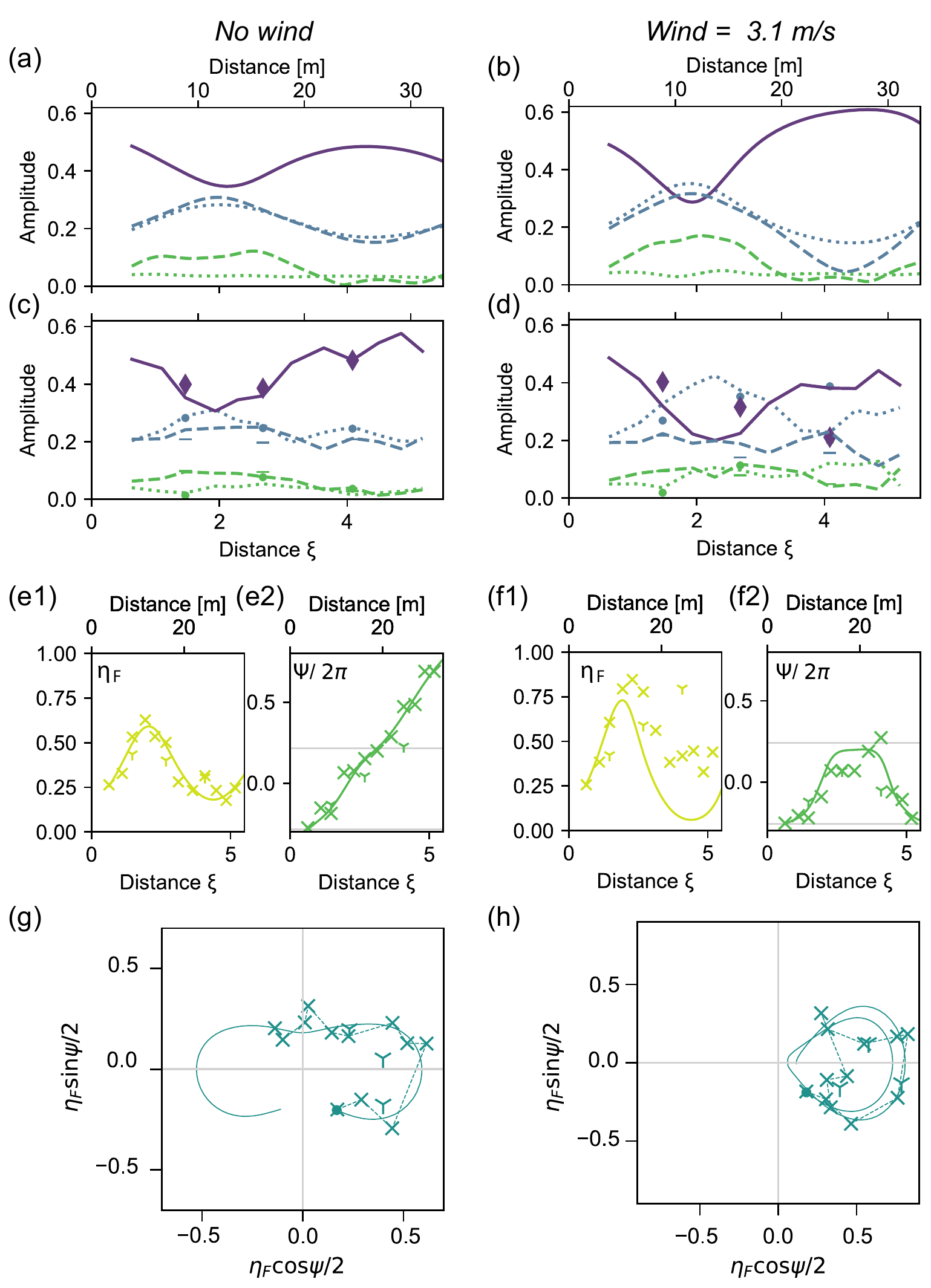}
\caption{a,b,c,d) Evolution of the 5 main modes of the spectrum: a) Simulation: Wind = 0 m/s, b) Simulation: Wind = 3.1 m/s, c) Experiment: Wind = 0 m/s, d) Experiment: Wind = 3.1 m/s. Simulation and measurement of $\eta_\text{F}$ without wind (e1) and Wind = 3.1 m/s (f1). Simulation and measurement of $\psi$ without wind (e2) and Wind = 3.1 m/s (f2). Simulated and measured evolution in the phase-space spanned by $\eta_F$ and $\psi$ without wind (g) and Wind = 3.1 m/s(h)} \label{fig:modesPhase}
\end{figure*}

\Cref{fig:modesPhase}a shows the simulated mode evolution without wind. The standard FPUT pattern unfolds: the main mode decreases and the sidebands reach a maximum at the focal point ($\xi=2$,~$x=12$ m). After the focus, the modulation is reversed.  This pattern is qualitatively the same in the experiments (\cref{fig:modesPhase}b) where however the lower sideband is more dominant. This downshift of the peak can be attributed to the transverse modes \cite{Trulsen1999}. The system crosses the vertical axis of the phase plane for both simulation and experiment (\cref{fig:modesPhase}g), corresponding to the phase shift observed in the real space in \cref{fig:envelope}a,c. This a consequence of the monotonically increasing trend of the phase $\psi$, displayed in the inset of \cref{fig:modesPhase}e2, giving a vertical-crossing every time  $\psi$ crosses $n\pi$. The sideband fraction $\eta_F$ (\cref{fig:modesPhase}e1) determines radial coordinate distance in the phase-space. 

With wind blowing at 3.1 m/s  the unbalance observed between the sidebands, as well as its underestimation by the simulations, are increased (\cref{fig:modesPhase}b,d). As this movement was already there in the case without wind, it is amplified by the presence of wind. The phase evolution in \cref{fig:modesPhase}f2 shows that the phase decreases before the $\pi$ limit, avoiding the crossing of the vertical axis in the phase plot (h), confirming that we are dealing with P1 solutions. As the initial condition was outside the separatrix (P2 regime), the crossing of the separatrix occurred during the evolution.

An alternative approach to achieve separatrix crossing during the propagation can be achieved by propagating waves over an smoothly increasing finite depth \cite{Armaroli2020_smooth}.

\subsection{\label{sec:Wavebreaking}Balance between dissipation and forcing}
\begin{figure*}
\centering
\includegraphics[width=0.8\textwidth]{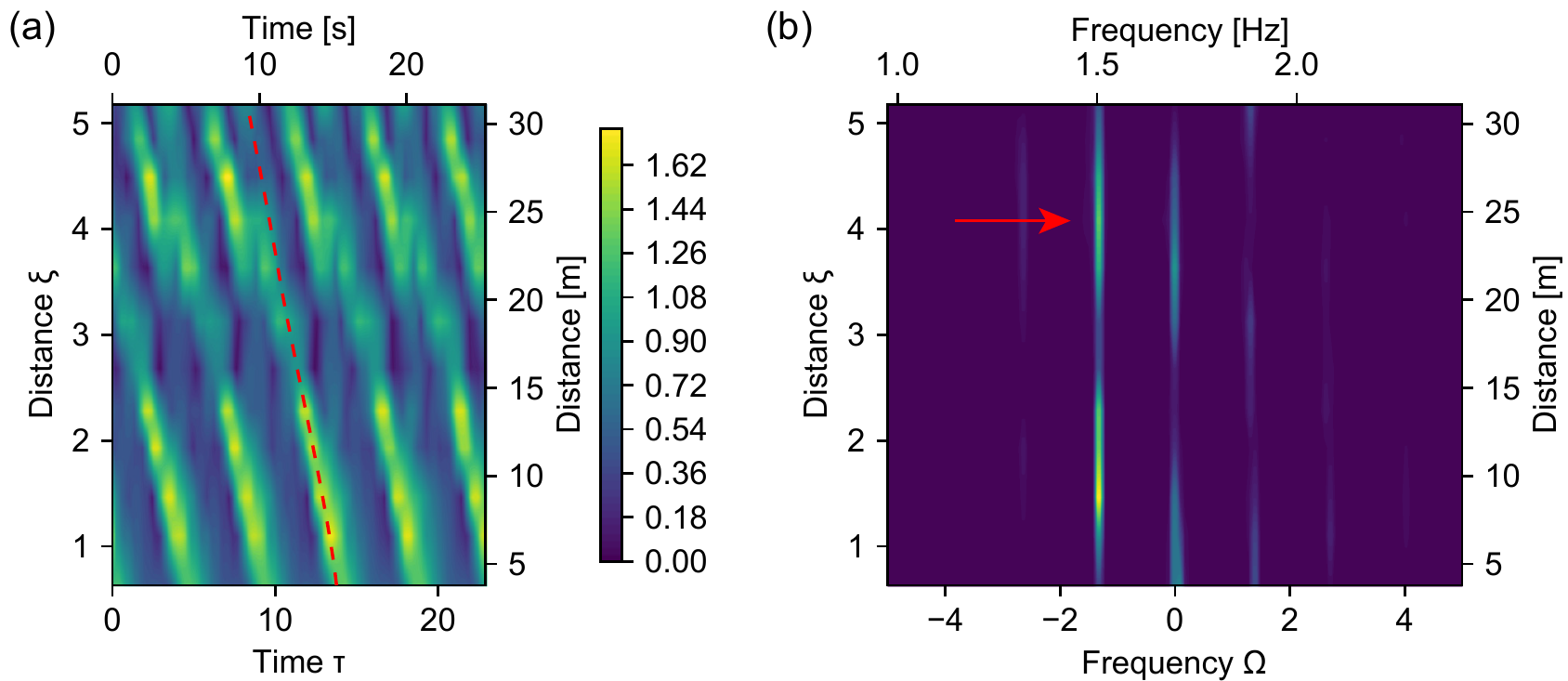}
\caption{a) Envelope evolution at strong wind forcing (4.0 m/s),  showing a phase-shift. b) Corresponding spectral evolution. The lower sideband stays dominant after the first recurrence cycle. This permanent downshift is indicative of wave breaking, and consequently dissipation of the breather.} \label{fig:waveBreaking}
\end{figure*} 
If the wind is not sufficient to outbalance the viscous damping the solution will not cross the separatrix. With our specific initial condition, dissipation stayed dominant for wind speeds below 2.5 m/s. On the other hand, if the wind is too strong, as for $W$ = 4.0 m/s in \cref{fig:waveBreaking}, wave breaking will occur, which is a form of dissipation. 

In an FPUT recurrence cycle, the maximum modulation at the focal point coincides with the maximum spectral width. After focusing, the initial main mode will become dominant again. However, if the initial steepness is too high, or wind forcing too strong, wave breaking will cause a permanent downshift to the lower sideband. \Cref{fig:waveBreaking}a shows the resulting phase-shift of the envelope. This might also be the start of soliton fission, which occurs when two sidebands are within the MI range~\cite{Kimmoun2017}. Wind forcing expands the MI-range, allowing the second mode $2\Omega_{\textrm{mod}}=2.82$ to lie within the unstable range. From both observation by eye and the permanent downshift shown in \Cref{fig:waveBreaking}b, we can conclude that a wave breaking event has indeed occurred.

\subsection{\label{sec:outsideMI}Behavior outside the MI-band}
\begin{figure*}
\centering
\includegraphics[width=0.7\textwidth]{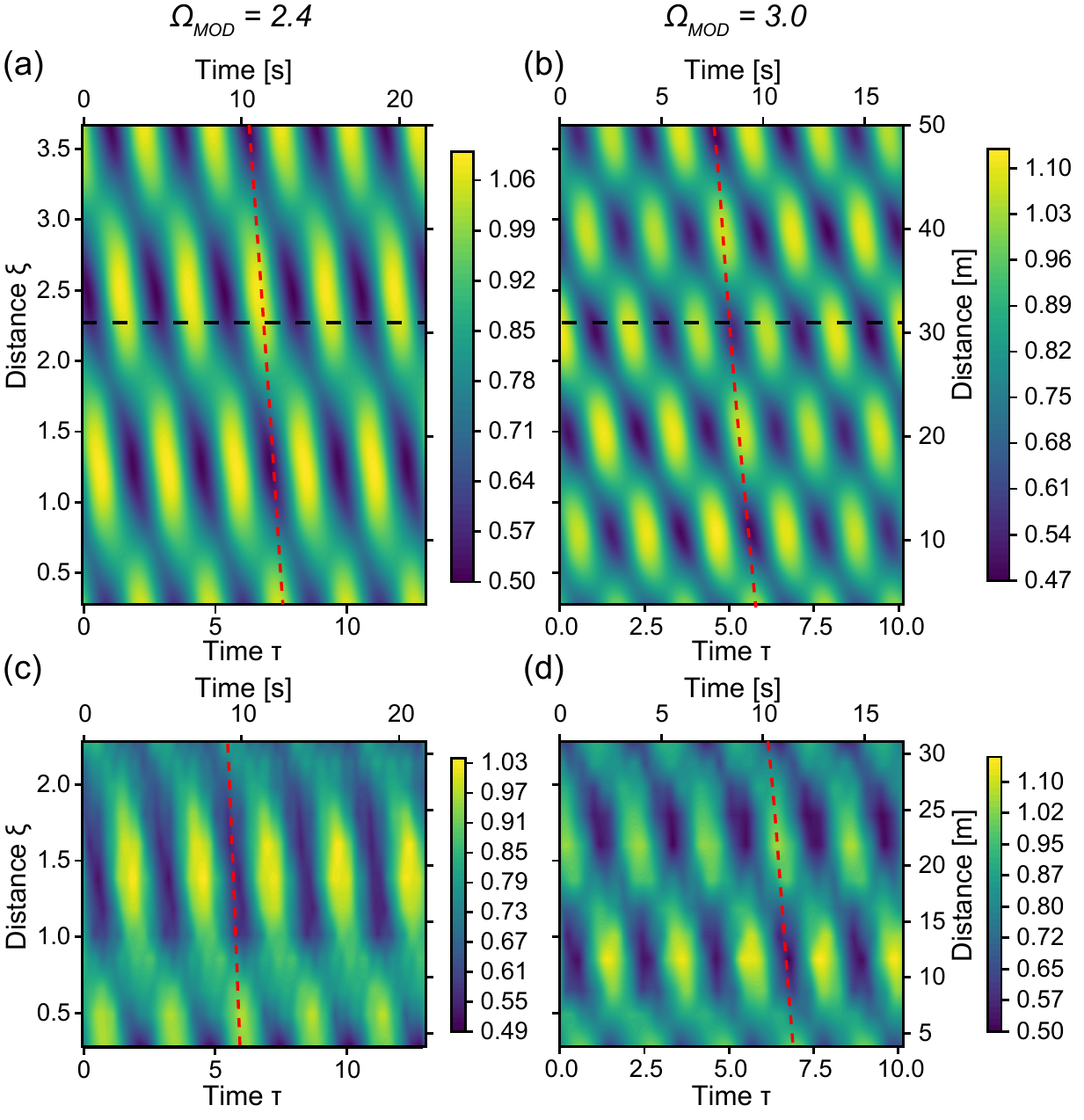}
\caption{Phase-shifted evolution of the envelope of a harmonically perturbed plane wave without wind, for $\Omega_{\text{mod}}$ outside the MI-band. Wave properties: $f_0 = 1.35$ Hz, $\alpha=0$, $\eta_F$=0.05. The horizontal dashed line indicates physical tank limit. a) Simulation: $\Omega_{\text{mod}} = 2.4$ b) Simulation: $\Omega_{\text{mod}} = 3.0$,  c) Experiment: $\Omega_{\text{mod}} = 2.4$, d) Experiment: $\Omega_{\text{mod}} = 3.0$. } \label{fig:outsideMI}
\end{figure*} 

We examined the evolution of initial conditions where $\Omega_{\text{mod}}$ is outside the MI-gain band. While the linear stability analysis predicts no growth of the modulation outside this limit, \cref{fig:PhaseSpace}a shows an oscillation of $\eta_\text{F}$ and thus of the amplitude for $\Omega_{\text{mod}} > 2$. A similar growth is predicted for Type A solutions of the conservative NLSE in~\cite{Conforti2019}.

The measured envelope evolution of the HPPW for $\Omega_{\text{mod}} = 2.4$ (\cref{fig:outsideMI}c)  and $\Omega_{\text{mod}} =3$ (\cref{fig:outsideMI}d) confirm the existence of this growth and decay cycle. Panels (a) and (b) display the corresponding simulations based on \cref{eq:Dysthe:FDadim}.  Following the dashed line of the group velocity, we indeed also observe the expected phase-shift of the envelope as only P2 solutions exist in this regime (see \cref{fig:PhaseSpace}a). In addition, comparing the different modulation frequencies shows that the spatial recurrence period (in $\xi$) is inversely proportional to the temporal modulation period $\Omega_{\text{mod}}$, as theorized in~\cite{Conforti2019}. This experimental result provides an important sanity check for the model. In addition, despite the spectral asymmetries and viscous dissipation present in the water waves, we still qualitatively observe the same behavior as a solution to the `pure' conservative NLSE.
\begin{figure*}
\centering
\includegraphics[width=0.9\textwidth]{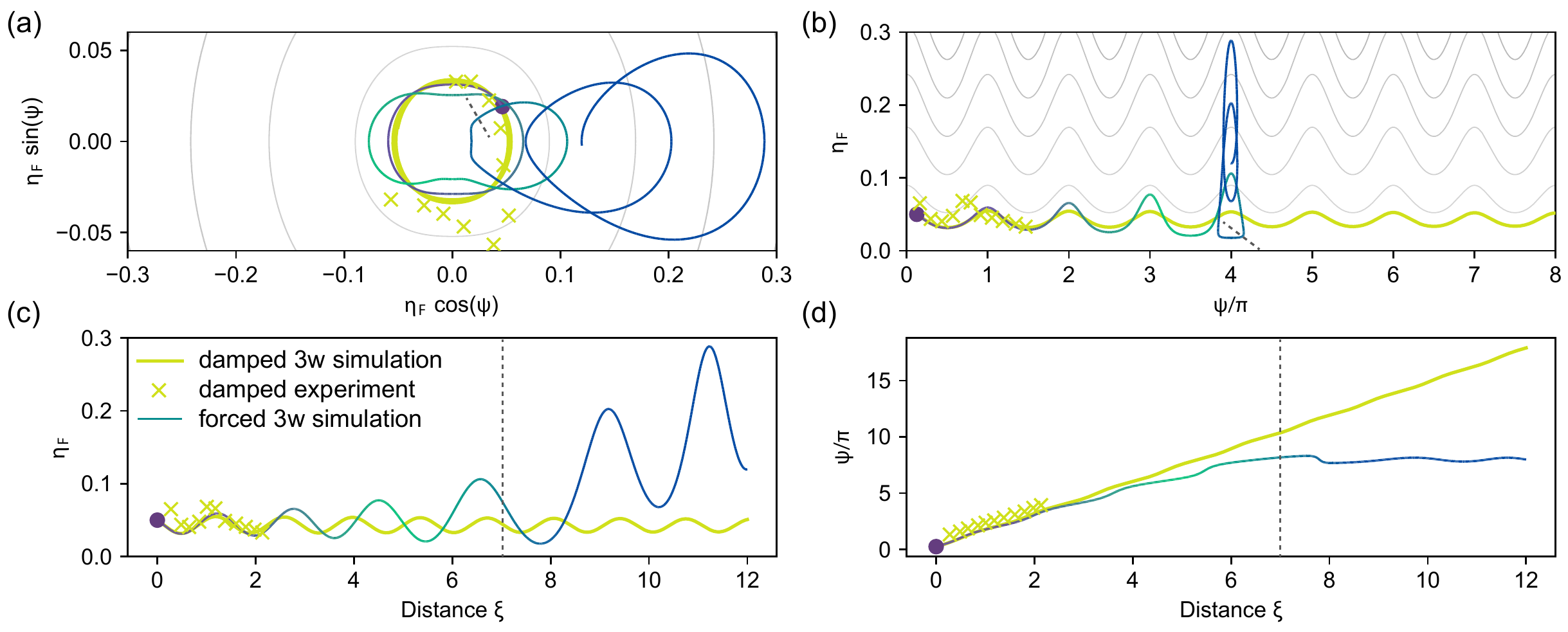}
\caption{Three-wave simulated evolution of damped wave outside the MI band (yellow line), and of a forced wave (line colored from purple to green to blue for increasing distance). The experimental measurements of the damped wave that correspond to \cref{fig:outsideMI}c are indicated by the yellow crosses.  a) Wrapped phase space. b) Unwrapped phase-space. c) Evolution of sideband-fraction $\eta_F$. d) Evolution of relative phase $\psi$. The dashed line indicates the separatrix crossing} \label{fig:outsideMI_sim}
\end{figure*} 

We now ask ourselves what happens if we include forcing. \Cref{fig:outsideMI_sim} shows the result for the three-wave simulations for the same initial conditions as \cref{fig:outsideMI}a, undergoing damping ($\delta_0=-0.006$), indicated by the yellow line, and forcing ($\delta_0=0.05$) indicated by the colored line. The yellow crosses show the measurements. For the length of the tank ($\xi \leq 2.3$), the forced and damped trajectory are nearly identical. Propagation length, wave breaking and transverse modes make deep water waves an impractical system to observe long-term or strong forcing. We therefore turn to simulations to examine the general behaviour of a forced NLS-system, not limited to water waves.     
 
The forced trajectory in \cref{fig:outsideMI_sim} shows that forcing the system can induce a symmetry breaking \textit{during} the evolution, that is, reshaping the potential landscape from a single well (\cref{fig:PhaseSpace}a1) to a double well (\cref{fig:PhaseSpace}b1), by moving ${\Omega_\text{mod}}$ from outside to inside the MI-band.

The trajectory undergoes three stages, indicated by different colors  i) Purple: the initial condition is in a single-potential well.  ii) Green: forcing increases the width of the MI-band, such that during the evolution ${\Omega_\text{mod}}$ will move inside the MI-band, creating a double-well potential, but the trajectory will still be outside of the separatrix (P2).  iii) Blue:  in the double potential well landscape, the trajectory is attracted to the P1 solution due to forcing. After crossing the separatrix it therefore remains in one lobe in \cref{fig:outsideMI_sim}a. Indeed, \cref{fig:outsideMI_sim}d shows that $\psi$ increases until ${\Omega_\text{mod}}$ enters inside the separatrix at  $\xi = 7$, as indicated by the dashed line, and $\psi$ starts to oscillate within a range of width $\pi$. As the behaviour of the trajectory in stages (i) and (ii) is qualitatively the same, the phase-locked behavior in stage (iii) shows that indeed a double well potential is formed and thus symmetry breaking has occurred.

We note that a physical counterpart to this long-duration, linear forcing is challenging to find, as the forcing will likely become saturated or nonlinear. Nevertheless, this simulation demonstrates the drastic effect of the expansion of the MI-band due to forcing on the NLSE behavior.

\section{Conclusion} 
We experimentally and theoretically examine the effect of forcing and damping on the recurrence cycle of NLSE-type solutions. Deep-water waves are well suited for our experimental study, as this system can be both forced (by wind) and damped (by viscosity). We distinguish between the cases where modulation frequency $\Omega_{\text{mod}}$ is inside and outside the MI-band.

We contribute five novel findings. 1) Our main finding is the experimental demonstration that when wind forcing is sufficient to overcome the viscous damping, the system is attracted toward P1 solutions, inducing a separatrix crossing during the evolution. We demonstrate such P1 behavior in the real space, by showing that the P2 phase-shift of the envelope is lifted. 2) We are able to reconstruct the phase-space trajectory from experiments using the three-wave truncation, allowing us to demonstrate the difference between P1 and P2 trajectories. 3) We show that if the wind forcing is too strong, it induces wave breaking, which is a form of energy dissipation that restores the P2 phase-shift. 4) We experimentally show that while no growth is expected outside the MI-band based on a linear stability analysis, there is in fact a growth and decay pattern of P2 solutions here, confirming the theoretical findings in Ref. \cite{Conforti2019}. 5) We theoretically show that forcing the system can induce symmetry breaking during the evolution, by moving $\Omega_{\text{mod}}$ from outside to inside the MI-band.  

Since NLSE-type solutions are found in systems other than water waves, we expect that our findings will be confirmed in different experimental setups under the effect of positive and negative forcing.

\begin{acknowledgements}We acknowledge the financial support from the Swiss National Science Foundation (Project No. 200021-
155970). We thank Alexis Gomel for fruitful discussions. We grateful to Olivier Kimmoun and Fabien Remy from IRPHE/ECM and Centrale Innovation for the use of their wave gauges. 
\end{acknowledgements}

\bibliography{PhaseShift}  

\end{document}